\documentstyle[12pt]{article}

 \newcommand{\be}{\begin{equation}}
 \newcommand{\ee}{\end{equation}}
 \newcommand{\ba}{\begin{eqnarray}}
 \newcommand{\ea}{\end{eqnarray}}
 
 \newcommand{\del}{\partial}

\newcommand{\lef}{\left}

\newcommand{\ri}{\right}

\newcommand{\cl}{{\cal L}}

\newcommand{\fr}{\frac}

\newcommand{\rmi}{\rm i}

\begin{document}

\begin{titlepage}

\topmargin -15mm

\vskip 10mm

\centerline{ \LARGE\bf Quantum Global Vortex Strings} \vskip 2mm
\centerline{ \LARGE\bf in a Background Field}

    \vskip 2.0cm

\centerline{\sc E.C.Marino }

\vskip 0.6cm

\centerline{{\it Instituto de F\'\i sica }} \centerline{\it
Universidade Federal do Rio de Janeiro } \centerline{\it Cx.P.
68528, Rio de Janeiro RJ 21941-972} \centerline{\it Brazil}

\vskip 1.0cm

\begin{abstract}

We consider quantum global vortex string correlation functions,
within the Kalb-Ramond framework, in the presence of a background
field-strength tensor and investigate the conditions under which
this yields a nontrivial contribution to those correlation functions.
We show that a background
field must be supplemented to the Kalb-Ramond theory, in order to
correctly describe the quantum properties of the vortex strings. The explicit
form of this background field and the associated quantum vortex
string correlation function are derived. The complete expression for
the quantum vortex creation operator is explicitly obtained. We
discuss the potential applicability of our results in the physics of
superfluids and rotating Bose-Einstein condensates.

\end{abstract}

\vskip 2cm


\vskip 5 mm


\vskip 2cm
 Work supported in part by
CNPq and FAPERJ

\end{titlepage}

\hoffset= -10mm

\leftmargin 23mm

\topmargin -8mm
\hsize 153mm

\baselineskip 7mm
\setcounter{page}{2}


Global vortex strings appear as nontrivial excitations in many
important physical systems described by the relativistic global
U(1) Abelian Higgs model in the spontaneously broken phase.
Typical examples are the cosmic strings left as remnants of topological
phase transitions occurred in the early universe \cite{twk}. Superfluid
vortices are similar excitations occurring in nonrelativistic systems
such as superfluid helium II \cite{don}.
The second case is evidently a completely quantum system and, therefore a
full quantum description of the superfluid vortex excitations is
necessary. Even for cosmic strings, however, a quantum description
is likely to be needed in the early stages of the universe, where
quantum effects should be important.
Quite interesting works have been devoted to the quantum
description of vortices in superfluids \cite{qvd}. The particular case
of vortices in superfluid films has been considered in \cite{qftsfs} where
a topological coupling between the field associated to the superfluid film
and the vortices has been introduced. Yet, a formulation made explicitly in terms of a
fully quantized vortex creation operator and its correlation functions
would be highly interesting.

More recently, it has been found
that superfluid vortices are abundantly generated in rotating
Bose-Einstein condensates. These are also deeply quantum systems and
consequently a quantized framework is required for studying the
dynamics of such vortices. An intense research activity has been
devoted to these systems in the recent years \cite{vbec}.

The relation between global strings and superfluid vortices has been
clarified quite a few years ago. It has been shown that classical
superfluid vortices could be described as global strings in the
presence of a particular nonrelativistic background field
\cite{ds,hyat}. Assuming that this relationship could be extended up
to the quantum level, the potential applicability of a quantum
theory of global strings to superfluid vortices becomes clear.

In a previous paper, we introduced a creation operator of fully
quantized global vortex string states and evaluated its two-point
correlation function in the Kalb-Ramond framework \cite{fm}. As we
argue below, however, this correlation function is not satisfactory.
The reason is its short distance behavior does not correspond to
what one should expect from a genuine operator creating local
states. The fact that these are not normalizable implies that the
corresponding correlator should diverge at short distances.

In this work, we evaluate the global vortex string creation operator correlation
function in the presence of a background field and investigate the
necessary conditions for a nontrivial effect thereof. We show that a
background field containing the information that the superfluid
state is destroyed along the vortex string should be included, in order for
the quantum correlation function to have the nontrivial short
distance behavior suitable to a genuine operator creating local
quantum states. The vortex string correlation functions in the presence of
a background are equivalent to the correlation functions of a new
vortex string operator, whose form is explicitly obtained.

Global vortex strings are solutions of theories with spontaneously
broken global U(1) symmetries. One of the simplest
examples is the U(1) Higgs model,
\be \cl[\phi] = |\partial_\mu
\phi|^2 + m^2 |\phi|^2 - \fr{\lambda}{4} |\phi|^4, \label{1}
\ee
where $\phi =\fr{\rho} {\sqrt{2}}e^{i\theta}$ is a complex scalar
field. Another example would be the Gross-Pitaevskii equation,
which is nonrelativistic \cite{gp}.

An interesting particular regime is the one when the coupling
$\lambda$ is large and we may approximate $\rho$ by its constant
vacuum value $\rho_0= 2m^2/\lambda$. The only remaining dynamical
degree of freedom, in this case, is the scalar (Goldstone boson)
field $\theta$ \be \cl = \frac{\rho_0^2}{2} \partial_\mu \theta \partial^\mu
\theta. \label{11} \ee

In this framework, global strings are solutions with a nonzero
vorticity. This is defined in terms of the vorticity current \be
J^{\mu \nu}(x) \equiv \epsilon^{\mu \nu \rho \sigma}\partial_\rho
\partial_\sigma \theta(x) = 2\pi n \int_{S(C)} d^2 \xi^{\mu\nu}
\delta^4(x-\xi) ,\label{2} \ee where the vortex string coincides
with the curve $C$ and $S(C)$ is its world-surface. In the above
equation, for the vortex current to be nonvanishing, of course,
$\theta$ must be multivalued. $2\pi$ is the vorticity quantum in the
units we are working and $n$ is an integer corresponding to the
number of vortex quanta. Assuming that the phase $\theta$ is defined
relatively to the vortex center, the second part of (\ref{2})
follows.

An extremely useful equivalent description of this system was
obtained \cite{sfkr}, in terms of the antisymmetric tensor gauge
field $B_{\mu\nu}$, or Kalb-Ramond field, whose field-strength
tensor is given by $H_{\alpha\beta\gamma}=\partial_\alpha B_{\beta
\gamma} +
\partial_\beta B_{\gamma \alpha} + \partial_\gamma B_{\alpha \beta}$ \cite{kr}.
The connection is made by writing the U(1) Higgs current as the
topological current of the Kalb-Ramond field, namely, \be J_\mu =
\fr{1}{6} \epsilon_{\mu\nu\alpha\beta}H^{\nu\alpha\beta}=
\rho^2\partial_\mu \theta. \label{3} \ee

It has been shown that in the regime where the field $\rho$ has a
constant value $\rho_0$, the Higgs lagrangian (1), in the presence
of a string configuration associated to the current (\ref{2}), can
be described by \cite{sfkr}
\be \cl[B_{\mu \nu}]= \fr{1}{12 \rho_0^2
} H_{\mu\nu\alpha}^2 + \fr{1}{2} B_{\mu\nu}J^{\mu \nu}. \label{4}
\ee

The vorticity flux along a surface $R$, is then given by (from now
on, we assume that the $i,j,k$-components run over spatial indexes)
\be \Phi_R = \int_R d^2 x^i J^{i0} (\vec x,t) = \int_R d^2 x^i
\partial_j \Pi^{ji}  , \label{22} \ee where $\Pi^{ij}$ is the
momentum canonically conjugate to the Kalb-Ramond field.

An important application of this theory is in the description of
superfluid systems. In these, the superfluid density is given by
$\rho^2$ and the superfluid velocity, by $\vec \nabla\theta$. The
superfluid current, therefore, is $\vec j = \rho^2 \vec
\nabla\theta$. This is the spatial component of $j_\mu = \rho^2
\partial_\mu \theta = - \rmi \phi^* \stackrel{\leftrightarrow}{\del}_\mu
\phi$, namely, the U(1) current of the Higgs model, (\ref{1}). The
following point, however, is crucial for using the Kalb-Ramond field in
the description of the superfluid.
It has been demonstrated that in order to correctly describe the classical
physical properties of vortices in superfluid systems, one
should add to the field-strength tensor $H^{\mu\nu\alpha}$ a
constant non-relativistic background field \cite{ds,hyat}
\be
\bar{H}^{\mu\nu\alpha} =
 \lef \{  \begin{array}{c}
\sqrt{\rho_0}\   \epsilon^{ijk} \ \ \ \ \ \ \ \ \ \ \ \ \ \ \ \
\mu,\nu,\alpha =i,j,k
  \\    \\
 0 \ \ \ \ \ \ \ \ \ \ \ \ \ \ \ \ \ \ \ \ \ \ \  {\rm otherwise}
          \end{array} \ri .
\label{5} \ee
The reason why global strings differ from superfluid vortices is that
they live in a Lorentz invariant vacuum and, consequently, there is no
circulation of energy-momentum around them. Superfluid vortices, conversely,
do have a fluid flux circulating them. It has been shown
however, that in the presence of the nonrelativistic background
(\ref{5}) the global string spins around its axis and superfluid vortices
are equivalent to spinning global strings \cite{ds,hyat}.
This description is an alternative to the Gross-Pitaevskii formulation
that would lead to the nonrelativistic version of (\ref{11})
in the (constant $\rho$) incompressible regime (see (\ref{20})).

In a previous paper \cite{fm}, using a general method of
quantization of topological excitations \cite{ev}, we have
introduced a quantum global string creation operator, $\sigma(C,t)$,
which acting on the vacuum, creates a quantum string state at the
curve $C$ on the instant $t$. It is given by \be \sigma(C,t) = \exp
\lef \{ \fr{ia}{2} \int_{T(C)} d^2 \xi_{\mu\nu} \fr{\del_\alpha
H^{\alpha\mu\nu}}{-\Box} \ri \}= \exp \lef \{ \fr{-ia}{2}
\int_{T(C)} d^2 \xi_{ij} B^{ij} + gauge\ terms \ri \}.
 \label{6} \ee
In the above expressions, $a$ is an arbitrary real number and
$T(C)$ is a space-like surface bounded by the closed string at
$C$. $d^2 \xi_{ij}$ is the surface element of $T(C)$, the
directions $i,j$ being along the surface.

We may write (\ref{6}) as \be \sigma(C,t) = \exp \lef \{ \fr{1}{2}
\int d^4 x H_{\alpha\mu\nu}\tilde{C}^{\alpha\mu\nu} \ri \},
\label{7} \ee where \be
\tilde{C}^{\alpha\mu\nu}=\partial^\alpha\tilde{C}^{\mu\nu} \ \ \ \
;\ \ \ \  \tilde{C}^{\mu\nu}=ia\int_{T(C)} d^2
\xi_{\mu\nu}\fr{1}{-\Box} (z-\xi). \label{8} \ee

From (\ref{22}) and (\ref{6}), we can show \cite{fm} that  \be
\Phi_R |\sigma(C,t)\rangle = a \ |\sigma(C,t)\rangle, \label{9} \ee
provided $C$ pierces $R$. This demonstrates that indeed the quantum
string operator creates eigenstates of the vorticity operator with
eigenvalue $a$, which we may choose equal to $2\pi n$ (notice that
$\Phi_R$ and also $a$ are dimensionless).

The euclidean two-point correlation function of the quantum string
operator can be written as the functional integral \cite{fm}
$$
\langle\sigma (C_x)\sigma ^{\dagger}(C_y)\rangle=Z^{-1} \int
DB_{\mu \nu} \exp\left \{-\int d^{4}z \left [\fr{1}{12 \rho_0^2}
H^{\mu\nu\alpha} H_{\mu\nu\alpha} + \right. \right.
$$
\be \left. \left. +\fr{1}{6}\tilde{C}_{\mu \nu \alpha}(z;x,y)
H^{\mu \nu \alpha} + \fr{\rho_0^2}{6}\tilde{C}^{\mu \nu \alpha}
\tilde{C}_{\mu \nu \alpha}
 \right ]  \right \},
\label{9} \ee where $Z$ is the vacuum functional,
$\tilde{C}_{\mu \nu \alpha}(z;x,y) =
\tilde{C}_{\mu \nu \alpha}(z;x) - \tilde{C}_{\mu \nu
\alpha}(z;y)$, corresponding to strings in $C_x$ and $C_y$. The
last term in the exponent is a renormalization counterterm,
introduced in order to guarantee locality of the correlation
function, which thereby becomes surface independent. An arbitrary
n-point quantum global string correlation function would be
obtained by just inserting additional external fields
$\tilde{C}_{\mu \nu \alpha}(z;x_i), \ i=1...n$ in (\ref{9}).

In \cite{fm}, we have calculated the above correlator, at equal
times, in the case of a large straight string of length $L$ along
the $z$-direction and piercing the $xy$-plane at the point $\vec x$.
The result is

\be \langle\sigma(\vec x,t)\sigma^\dagger(\vec y,t)\rangle = \exp
\left
 \{-\fr{La^2\rho_0^2}{8\pi}|\vec x-\vec y|
\right \} .\label{10} \ee From the large distance behavior of this
expression, we can obtain the quantum string creation energy. This
is given by $E(L)=\fr{La^2\rho_0^2}{8\pi}$ which means that the
string energy per unit length is a constant parametrized by the
vorticity of the quantum string and the superfluid density.

Even though the large distance behavior of expression (\ref{10}) is
suitable and reveals the energy of the objects being created by
$\sigma$, its short distance is unsatisfactory. Observe that \be
\lim_{\vec x \rightarrow \vec y}\ \ \langle\sigma(\vec
x,t)\sigma^\dagger(\vec y,t)\rangle = \ |\ |\sigma(\vec x,t)\rangle\
|^2 = 1 .\label{11} \ee Remember, however, that genuine local states
created by a quantized field are not normalizable. Hence, the two
point correlator of a genuine quantum global string creation
operator should diverge at short distances. In what follows, we show
that this shortcoming can be eliminated by the introduction of a
particular background Kalb-Ramond field strength tensor in
(\ref{9}). By the way, as we have seen before, the correct classical
description of superfluid vortices also requires the inclusion of a
background field given by (\ref{5}).

Following the above reasoning, instead of (\ref{9}), we are going
to evaluate
$$
\langle\sigma (C_x)\sigma ^{\dagger}(C_y)\rangle_{\bar H}=Z^{-1}
\int DB_{\mu \nu} \exp\left \{-\int d^{4}z \left [\fr{1}{12
\rho_0^2} \left(H^{\mu\nu\alpha} + \bar
H^{\mu\nu\alpha}\right)\left( H_{\mu\nu\alpha} + \bar
H^{\mu\nu\alpha}\right) + \right. \right.
$$
\be \left. \left. +\fr{1}{6}\tilde{C}_{\mu \nu \alpha}(z;x,y)
\left(H^{\mu \nu \alpha}  + \bar H^{\mu\nu\alpha} \right)+
\fr{\rho_0^2}{6}\tilde{C}^{\mu \nu \alpha} \tilde{C}_{\mu \nu
\alpha}
 \right ]  \right \},
\label{12} \ee where the background field strength $\bar
H^{\mu\nu\alpha}$ is {\it a priori} non specified.

The functional integral over the $B_{\mu \nu}$-field is quadratic
and can be performed by following the same steps used for evaluating
(\ref{9}) \cite{fm}. The result is
$$
\langle\sigma (C_x) \sigma ^{\dagger}(C_y)\rangle_{\bar H} = \exp
\left \{ F_{\bar H =0}\left(C_x, C_y \right) - \frac{\rho^2_0}{2}
\int d^4z d^4z'\right .
$$
$$  \times \left\{\left[\frac{1}{6\rho_0^2} \epsilon^{\alpha \beta \gamma
\sigma}
\partial_\sigma \bar H_{\alpha\beta\gamma}(z)\right]
\left[ \frac{1}{6\rho_0^2} \epsilon^{\mu \nu \lambda \rho}
\partial'_\rho \bar H_{\mu\nu\lambda}(z') \right  ] + \left[
\epsilon^{\alpha \beta \gamma \sigma}
\partial_\sigma \tilde{C}^{\alpha\beta\gamma}(z)\right]
\left[ \frac{1}{6\rho_0^2} \epsilon^{\mu \nu \lambda \rho}
\partial'_\rho \bar H_{\mu\nu\lambda}(z') \right  ]\right\}
$$
\be \left . \times\left [ \frac{1}{-\Box}\right ](z-z') \right \}.
\label{13} \ee In this expression, $F_{\bar H =0}\left(C_x, C_y
\right)$ is the result of the calculation without any background,
performed in \cite{fm} and \be \left [ \frac{1}{-\Box}\right ](x)=
\int\frac{d^4k}{(2\pi)^4}\ \frac{e^{{\rm i k\cdot x}}}{k^2} =
\frac{1}{4\pi^2 |\vec x|^2} \ee

The second term in the exponent in (\ref{13}) vanishes because
$\tilde{C}^{\alpha\beta\gamma}$ is a derivative (see eq. (\ref{8})).
Accordingly, for the first term to be different from zero, the
background field should {\it not} be of the form $\bar
H_{\alpha\beta\gamma}=\partial_\alpha \bar B_{\beta \gamma} +
\partial_\beta \bar B_{\gamma \alpha} + \partial_\gamma \bar B_{\alpha
\beta}$. In this case the first term in (\ref{13}) would also
vanish. It is easy to see that when the background has this form
it gives no contribution to (\ref{12}): by shifting the functional
integration variable as $B_{\mu\nu}\rightarrow B_{\mu\nu}+ \bar
B_{\mu\nu}$ we would eliminate the background from (\ref{12}). It
is also clear from (\ref{13}) that a constant background such as
(\ref{5}) would have no effect in the quantum string correlation
function.

The condition for the background field to give a nonzero
contribution to the correlator (\ref{12}) is that it should {\it
not} satisfy Bianchi's identity, namely,

\be \frac{1}{6} \epsilon^{\alpha \beta \gamma \sigma}
\partial_\sigma \bar H_{\alpha\beta\gamma} \neq 0.
\label{14}
\ee

According to (\ref{3}), however, this means that the superfluid
current would be no longer conserved in the presence of the
background $\bar H_{\alpha\beta\gamma}$. This observation provides
us the clue for obtaining a suitable form for the background field
$\bar H_{\alpha\beta\gamma}$. Along a vortex the superfluid phase
is destroyed and, consequently, the superfluid current is no
longer conserved because of depletion. The same thing should
happen for a quantum vortex string. Hence, when computing the
quantum string correlator (\ref{12}), we must introduce a
background field such that Bianchi's identity is not satisfied
along the vortices. The simplest configuration satisfying this
requirement is

\be
\frac{1}{6} \epsilon^{\alpha \beta \gamma \sigma}
\partial_\sigma \bar H_{\alpha\beta\gamma} = \rho_0\  a\
\oint_{C_x-C_y}d\xi^\mu\ \hat{n}_\mu(\xi)\ \delta^4(z-\xi),
\label{15} \ee where $\hat{n}_\mu(\xi)$ is a unit vector tangent to
the string at the point $\xi$. The background field strength
satisfying (\ref{15}) is

\be \bar H^{\alpha\beta\gamma} = \rho_0  a\
 \epsilon^{\alpha \beta \gamma \nu}
\int_{T(C_x)-T(C_y)}d^2\xi_{\mu\nu}\ \hat{n}^\mu(\xi)\
\delta^4(z-\xi), \label{16} \ee where $T(C)$ is a spacelike surface
bounded by $C$. Inserting (\ref{16}) in (\ref{13}) and performing the $z$ and
$z'$ integrals, we obtain the following expression for the ${\bar
H}$ contribution to the correlation function:

\be
 \exp \left \{
 \fr{ - a^{2}}{2}\sum_{i,j=1}^{2}\lambda_{i}\lambda_{j}
\oint_{C_i}d\xi^{\mu} \oint_{C_j}d\eta^{\nu}
\hat{n}_\mu(\xi)\hat{n}_\nu(\eta)
 \right. \lef. \lef [
\fr{1}{-\Box}  \ri ]( |\xi - \eta|  ) \ri \}. \label{17} \ee In this
expression, $\lambda_i =\pm 1$, corresponding to $C_x$ and $C_y$,
respectively. The self-interaction terms $i-i$ are eliminated by a
renormalization of the string operator $\sigma$.

Considering again the previous situation of a large straight string
along the $z$-axis and piercing the $xy$-plane at $\vec x$, we may
follow the procedure described in \cite{m96a} to evaluate
(\ref{17}). Including the $\bar H =0$ contribution, calculated
before, we get

\be \langle\sigma(\vec x,t)\sigma^\dagger(\vec y,t)\rangle_{\bar H}
= \exp \left
 \{\frac{L a^2}{|\vec x-\vec y|}-\fr{La^2\rho_0^2}{8\pi}|\vec x-\vec y|
\right \} .\label{18} \ee

We now have
\be \lim_{\vec x \rightarrow \vec y}\ \
\langle\sigma(\vec x,t)\sigma^\dagger(\vec y,t)\rangle = \ |\
|\sigma(\vec x,t)\rangle\ |^2 \rightarrow\infty,\label{19} \ee which
is the expected behavior of a genuine creation operator of local
states. The large distance behavior is the same as before, yielding
a string tension $\fr{a^2\rho_0^2}{8\pi}$.

In the same way that the Kalb-Ramond theory should be supplemented
by the addition of the background field (\ref{5}) in order to
correctly describe the classical properties of vortex strings
\cite{ds}, it must be supplemented by the introduction of the
background field (\ref{16}), in order to describe the quantum
properties of these excitations. It would be interesting to study
the behaviour of $\langle\sigma\rangle$, in order to investigate the
possible relationship between the two background fields.

An alternative interpretation that is quite appealing, is to regard
(\ref{12}) and (\ref{18}) as the correlator of a new vortex string
creation operator $\Sigma \equiv \sigma\mu$. From  (\ref{12}) and
(\ref{16}), dropping the renormalization terms, we find that the
operator $\mu$ is given by \be \mu(C,t) = \exp \lef \{
\fr{ia}{2\rho_0} \int_{T(C)} d^2 \xi_{\mu\nu}\hat{n}^\nu
\epsilon^{\mu\alpha\beta\gamma} \del_\alpha B_{\beta\gamma}\right \}
=
 \exp \lef \{ \fr{ia}{2\rho_0} \int_{T(C)}
d^2 \xi_{ij}\hat{n}^j \epsilon^{ikl} \Pi^{kl}  \ri \},
 \label{19} \ee
where $ \Pi^{ij}$ is the momentum canonically conjugate to the
Kalb-Ramond field $ B^{ij}$.

Eq. (\ref{9}) would be also satisfied by the states
$|\Sigma(C,t)\rangle$, since $\mu$ commutes with $\Phi_R$. This
shows that $\Sigma$ indeed is a quantum global string creation
operator. Notice that it resembles the Mandelstam operator of 2D
field theories \cite{mand}. This suggests a parallelism between the
systems studied here and the two-dimensional ones where the
Mandelstam operator is relevant, such as the sine-Gordon theory and
the Coulomb gas.

Expression (\ref{18}) for the vortex string correlation function, and the new
quantum vortex creation operator obtained thereof
should be relevant in the study of
the dynamics of quantum vortices, vortex lattice formation
and in the vortex nucleation problem in superfluid helium II
and in rotating Bose-Einstein condensates.

Let us remark, finally, that the relevant physics of superfluids is
described by the Gross-Pitaevskii equation \cite{gp}, which is of first
order in time. In this case, the constant $\rho$ regime of the fluid
would be described by the nonrelativistic version of (\ref{11}), namely
\be
\cl = \frac{\rho_0^2}{2} \left [ i  \theta \partial_0 \theta +
\partial_i \theta \partial_i \theta \right ] .
\label{20}
\ee
Accordingly, in order to construct the vortex operator corresponding to $\sigma$ or $\Sigma$
directly in the Gross-Pitaevskii framework,
we should consider the nonrelativistic version of (\ref{4}), (\ref{9}), (\ref{11})
and (\ref{12}). We intend to pursue this in a future publication.



\begin{thebibliography}{99}


\bibitem{twk} T.W.B. Kibble, {\it Phys. Rep.} {\bf 67} (1980) 183.

\bibitem{don} R.J. Donnelly, {\it Quantized Vortices in Helium II},
Cambridge Studies in Low Temperature Physics, Cambridge Univ. Press
1991 (Cambridge);
E.B.Sonin, {\it Rev. Mod. Phys. } {\bf 59 } (1987) 87;
M.Rasetti and T.Regge in Highlights of Condensed Matter Theory,
F.Bassani et al. eds. (Compositori, Bologna 1985);
V.Penna and M.Spera, {\it J. Math. Phys. } {\bf 30 } (1989) 2778


\bibitem{qvd} "Quantized Vortex Dynamics and Superfluid Turbulence",
Lecture Notes in Physics 571, C.F.Barenghi, R.J.Donnelly and W.F.Vinen,
editors, (Springer, Berlin 2001), p. 445

\bibitem{qftsfs} D.Arovas and J.Freire, {\it Phys. Rev.} {\bf B55 } (1997) 1068;
J.P.Kottmann and A.M.J.Schakel, {\it Phys. Lett.} {\bf A242 } (1998) 99

\bibitem{vbec} E.B.Sonin, {\it Phys. Rev.} {\bf A72 } (2005) 021606;
A.A.Aftalion, X.Blanc and J.Dalibard, {\it Phys. Rev.} {\bf A71 }
(2005) 023611; G.Baym, {\it Phys. Rev.} {\bf A69 } (2004) 043618;
V.Schweikhard, I.Coddington, P.Engels, V.P.Morgendorff and
E.Cornell, {\it Phys. Rev. Lett.} {\bf 92} (2004) 040404;
I.Coddington, P.Engels, V.Schweikhard and E.Cornell, {\it Phys. Rev.
Lett.} {\bf 91} (2003) 100402; J.Sinova, C.B.Hanna and
A.H.MacDonald, {\it Phys. Rev. Lett.} {\bf 89} (2002) 030403

\bibitem{gp} E.P.Gross, {\it N. Cimento} {\bf 6} (1961) 249; {\it J. Math. Phys. }
{\bf 4} (1963) 195;
L.P.Pitaevskii, {\it Zh. Eksp. Teor. Fiz.} {\bf 40} (1961) 646 [{\it Sov. Phys. JETP} {\bf 40} (1961) 451]

\bibitem{ds} R.L. Davis and E.P.S. Shellard, {\it Phys. Rev. Lett.}
{\bf 63} (1989) 2021.

\bibitem{hyat} M.Hatsuda, S.Yahikozawa, P.Ao and D.J.Thouless, {\it Phys. Rev.} {\bf B49}
(1994) 15870


\bibitem{fm} H.Fort and E.C.Marino, {\it Int. J. Mod. Phys.} {\bf A15} (2000)
2225

\bibitem{ev} E.C.Marino, {\it ``Dual Quantization of Solitons"}
in Proceedings of the NATO Advanced Study Institute {\it
``Applications of Statistical and Field Theory Methods to Condensed
Matter"}, D.Baeriswyl, A.Bishop and J.Carmelo , editors (Plenum,New
York)(1990)

\bibitem{sfkr} F.Lund and T.Regge, {\it Phys. Rev.} {\bf D14 } (1976)
1524; E.Witten, {\it Phys. Lett.} {\bf B153 } (1985) 243; A.Vilenkin
and T.Vachaspati, {\it Phys. Rev.} {\bf D4 } (1987) 1138; R.L.Davis
and E.P.S.Shellard, {\it Phys. Lett.} {\bf B214 } (1988) 219

\bibitem{kr} M.Kalb and P.Ramond, {\it Phys. Rev.} {\bf D9 } (1974)
2273

\bibitem{m96a} E. Marino, {\it Phys. Rev.} {\bf D53} (1996) 1001.

\bibitem{mand} S.Mandelstam, {\it Phys. Rev.} {\bf D11} (1975) 3026.



\end{thebibliography}
\end{document}